\documentclass[aps,prl, reprint, superscriptaddress]{revtex4-1}
\pdfoutput=1

\usepackage{graphicx}
\usepackage{amsmath}
\usepackage{amsfonts}
\usepackage{amssymb}
\usepackage{bm}
\usepackage{color}
\usepackage[normalem]{ulem}
\usepackage{soul} 
\usepackage{amssymb}
\usepackage{wasysym}
\usepackage{dsfont}
\usepackage{multirow}
\usepackage{epsfig}

\usepackage{bm}
\usepackage{amsmath,amsthm,amssymb}
\usepackage{array,booktabs,longtable}
\usepackage{mathrsfs}
\usepackage{subfigure}
\usepackage{color}
\usepackage{tikz}
\usetikzlibrary{decorations.markings}
\usetikzlibrary{calc,shapes,positioning,arrows,snakes}
\usepackage{array}
\usepackage{verbatim}
\usepackage{dsfont}
\usepackage{ifthen}
\newcommand{\bra}[1]{\langle #1 |}
\newcommand{\ket}[1]{| #1\rangle}

\DeclareMathOperator{\Tr}{Tr}

\newcommand{\bmm}{\begin{matrix}}
\newcommand{\emm}{\end{matrix}}
\makeatletter
\newcommand{\thickhline}{%
    \noalign {\ifnum 0=`}\fi \hrule height 1.25pt
    \futurelet \reserved@a \@xhline
}
\newcolumntype{"}{@{\hskip\tabcolsep\vrule width 1.25pt\hskip\tabcolsep}}
\makeatother

\begin{document}

% Use the \preprint command to place your local institutional report
% number in the upper righthand corner of the title page in preprint mode.
% Multiple \preprint commands are allowed.
% Use the 'preprintnumbers' class option to override journal defaults
% to display numbers if necessary
%\preprint{}

%Title of paper
\title{Entanglement dynamics in quantum many-body systems}

% repeat the \author .. \affiliation  etc. as needed
% \email, \thanks, \homepage, \altaffiliation all apply to the current
% author. Explanatory text should go in the []'s, actual e-mail
% address or url should go in the {}'s for \email and \homepage.
% Please use the appropriate macro foreach each type of information

% \affiliation command applies to all authors since the last
% \affiliation command. The \affiliation command should follow the
% other information
% \affiliation can be followed by \email, \homepage, \thanks as well.
%\author{}
%\email[]{Your e-mail address}
%\homepage[]{Your web page}
%\thanks{}
%\altaffiliation{}

\author{Wen Wei Ho}
\affiliation{Department of Theoretical Physics, University of Geneva, 24 quai Ernest-Ansermet, 1211 Geneva, Switzerland}
\affiliation{Perimeter Institute for Theoretical Physics, Waterloo, ON N2L 2Y5, Canada}

\author{Dmitry A. Abanin}
\affiliation{Department of Theoretical Physics, University of Geneva, 24 quai Ernest-Ansermet, 1211 Geneva, Switzerland}
\affiliation{Perimeter Institute for Theoretical Physics, Waterloo, ON N2L 2Y5, Canada}

%\date{\today}

\begin{abstract}
The dynamics of entanglement has recently been realized as a useful probe in studying ergodicity and its breakdown in quantum many-body systems. In this paper, we  study theoretically the growth of entanglement in quantum many-body systems and propose a method to measure it experimentally. We show that entanglement growth is related to the spreading of local operators in real space. We present a simple toy model for ergodic systems in which linear spreading of operators results in a universal, linear in time growth of entanglement for initial product states, in contrast with the logarithmic growth of entanglement in many-body localized (MBL) systems. Furthermore, we show that entanglement growth is directly related to the decay of the Loschmidt echo in a composite system comprised of several copies of the original system, in which connections are controlled by a quantum switch (two-level system). By measuring only the switch's dynamics, the growth of the R\'enyi entropies can be extracted. Our work provides a way of understanding entanglement dynamics in many-body systems, and to directly measure its growth in time via a single local measurement.
\end{abstract}

% insert suggested PACS numbers in braces on next line
% insert suggested keywords - APS authors don't need to do this
%\keywords{}

%\maketitle must follow title, authors, abstract, \pacs, and \keywords
\maketitle

% body of paper here - Use proper section commands
% References should be done using the \cite, \ref, and \label commands

\section{Introduction and results}
In the past decade, quantum entanglement has emerged as an indispensable tool for characterizing and classifying ground states of many-body systems, for example in the field of topological order~\cite{TO1,TO2}.
 %Further, in critical $(1+1)$-d many-body systems, the EE exhibits universal scaling with the subsystem size, with a prefactor determined by the central charge of the corresponding conformal field theory~\cite{CFT1, CFT3, CFT2}. 
%\textcolor{red}{Extraction of the central charge allows for an understanding of the universal behavior of the critical system.}
%The above examples refer to the static properties of entanglement in the ground states. 
%A natural question then arises: is the {\it dynamics} of entanglement universal, and is it a useful probe of different many-body phases? Indeed, this is so: 

Recently, it was realized that entanglement {\it dynamics} also exhibits universality, providing a useful probe of ergodicity and its breakdown in many-body systems. In particular, the growth of entanglement following a quantum quench can be used to distinguish between localized and ergodic phases. For many-body localized (MBL) systems~\cite{GMP, BAA}, emergent local integrability~\cite{AbaninSlowEE, MBLLIOM1, HuseSlowEE, MBLLIOM2} implies that entanglement grows logarithmically in time, $S(t) \sim \log t$, following a quantum quench from initial product states~\cite{MBLEEGrowth2,MBLEEGrowth1}. For generic quantum ergodic systems, entanglement appears to grow universally linearly~\cite{HuseFastEE}: $S(t) \sim t$. 

Previous works predicted linear growth $S(t) \sim t$ in integrable systems, including $(1+1)$-dimensional CFTs~\cite{CFTGrowth1,  EELinearGrowth}, while other works studied entanglement growth in non-integrable, higher dimensional CFTs via holographic calculations~\cite{HolographicGrowth1, HolographicGrowth2}. However, an understanding of the entanglement dynamics in generic quantum chaotic/ergodic systems is lacking. One puzzling fact pointed out in Ref.~\cite{HuseFastEE} is that energy transport in a quantum chaotic or ergodic many-body system under a time-independent Hamiltonian is diffusive ($\sim\sqrt{t}$), yet entanglement growth is linear ($\sim t$) -- clearly, the mechanisms of particle and quantum information transport are different, and so it is important to understand the dynamics of the latter. 

The purpose of this paper is to provide 1) a theoretical description, and hence, a physical picture of the growth of entanglement entropy (EE) in a many-body system, and 2) a proposal to experimentally measure it. To be precise,  we consider the dynamics of the $n$-th R\'enyi EE $S_n(t)$ of pure states $| \psi \rangle$ that are initially random product states, evolving unitarily under a local, potentially time-dependent non-integrable many-body Hamiltonian:
\begin{align}
H = \sum_X H_X.
\label{eqn:H}
\end{align}
The Hamiltonian is defined on a lattice $\Lambda$ of sites (labeled by $i$) in $d$ spatial dimensions so that the local Hilbert space $\mathcal{H}_i$ is bounded: $\text{dim}(\mathcal{H}_i) = k < \infty$, and $X$ is a local region in $\Lambda$. %The restriction that $H$ is local means that the size of the regions $|X|$ is uniformly bounded from above, for example, $2$ in the case of nearest neighbors. 
The $n$-th R\'enyi EE of $A$, for a bipartition of the system into two subregions $A$ and $B$, is given by
\begin{align}
S_n(t) = \frac{1}{1-n}\log \Tr(\rho^n_A(t)),
\end{align}
with $\rho_A(t) \equiv \Tr_{B}(U_t |\psi\rangle\langle \psi | U_t^\dagger)$ being the reduced density matrix of subsystem $A$, and $U_t$  the unitary time evolution operator generated by $H$ according to the Schrodinger equation. The state, being initially  a random product state, has $S_n(0) = 0$, but $S_n(t) > 0$ for $t > 0$ and grows in general. Also, $S_n \geq S_m$ for any $m > n$; in particular, $S_2$ is a lower bound for $S$, the von Neumann entropy.

A summary of our results are as follows. We show that the growth of entanglement as measured by $S_2(t)$ is directly related to the measurement of basis operators of subregion $A$. Under time evolution by a local Hamiltonian, a basis operator physically spreads in real space, and as it grows, the value of its measurement typically decreases, thus leading to entanglement growth. Furthermore, we introduce a simple toy model for ergodic many-body systems in which 1) operators spread at some maximal velocity $v$, 2) delocalize completely within this light cone $r \sim vt$, and show that $S_2(t)$ grows linearly in time with a velocity related to $v$. We believe such a model captures the salient features of the universal linear in time growth of entanglement seen by Ref.~\cite{HuseFastEE}.  

\begin{figure}[]
\center
\includegraphics[width=0.48\textwidth]{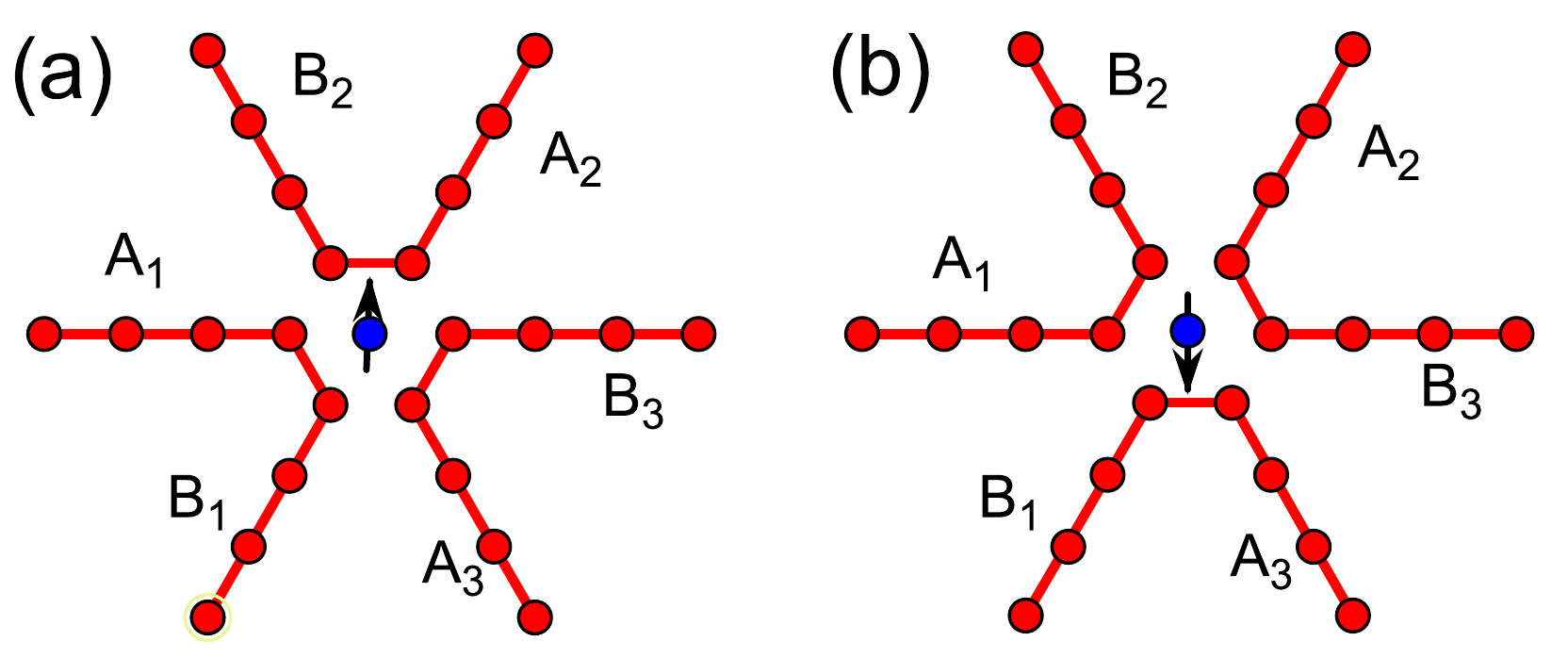}
\caption{Set-up to measure the growth of the $n$-th R\'enyi entropy $S_n(t)$. Here $n = 3$ and we have shown it for a $d = 1$ chain. There is a quantum switch in the middle of the set-up which governs tunneling between different subsystems. (a) When the quantum switch is in the state $\ket{\uparrow}$, tunneling between $A_i$ and $B_i$ is allowed while tunneling between $A_i$ and $B_{i+1}$ is prohibited. (b) When the quantum switch is in the state $\ket{\downarrow}$, the allowed and prohibited tunnelings are swapped. A composite state in an $n$-copy product state of the chains and in a superposition of $\ket{\uparrow}$ and $\ket{\downarrow}$ of the quantum switch will have two parts evolving in time differently according to the quantum switch; measuring $\sigma^x(t)$ of the quantum switch gives the Loschmidt echo and hence the $n$-th R\'enyi entropy. }
\label{fig:QS}
\end{figure}

We also propose a way to experimentally measure $S_n(t)$ via a local measurement. We introduce a quantum switch (a 2-level system) that allows tunneling between different parts of a replicated system consisting of $n$ disjoint copies of the original system, depending on the state of the switch~\cite{AbaninEE}. By preparing the replicated system and the quantum switch appropriately, and by subsequently measuring $\sigma^x(t)$ of the quantum switch only, the entropy growth $S_n(t)$ corresponding to the entanglement entropy of the original system can be measured, see Fig.~\ref{fig:QS}.

We will refine and expound upon these ideas below. For clarity of argument, we will focus on the case of a spin-$1/2$ system, so $\text{dim}(\mathcal{H}_i) = 2$, but our results are general and hold for systems with bounded local Hilbert spaces of other dimensions \cite{sup}. We also choose subregion $A$ to be a ball of radius $r_A$ with volume $\gamma_d r_A^d = N_A$ in $d$ dimensions ($\gamma_d = \frac{\pi^{d/2}}{\Gamma(d/2+1)}$ is the volume of a unit $d$-ball), but the analysis can be readily adapted to other geometries.  
%%%%%

\section{Local measurements of basis operators gives entanglement}
We first show that the second R\'enyi entropy $S_2(t)$ is directly related to measuring all time-evolved basis operators $\mathcal{O}_A (t)  = U_t^\dagger \mathcal{O}_A U_t $ in the initial state $\ket{\psi}$, such that the operator $\mathcal{O}_A$ at $t = 0$ has support strictly in $A$ (i.e. where it acts non-trivially on the lattice). That is, for a spin-$1/2$ system of $N_A$ sites, $S_2(t) = -\log \Tr_{A}(\rho_A^2(t))$, with
\begin{align}
\Tr_A\big(\rho_A^2(t) \big) = \frac{1}{2^{N_A}} \sum_{\mathcal{O}_A} \bra{\psi} \mathcal{O}_A(t) \ket{\psi} ^2,
\label{eqn:measureOperators}
\end{align}
where the sum goes over all operators $\mathcal{O}_A$ with support strictly in $A$, assumed to be Hermitian and independent as defined by the Hilbert-Schmidt inner product: $\frac{1}{2^{N_A}} \text{Tr}_A(\mathcal{O}_A^\dagger \mathcal{O}'_A)= \delta_{\mathcal{O}_A, \mathcal{O}'_A}$, and which form a basis for all Hermitian operators on $A$. A potential basis set is $\otimes_{i=1}^{N_A} \{ \mathbb{I}_i, \sigma_i^x, \sigma_i^y, \sigma_i^z \}$, where $\mathbb{I}_i, \sigma_i^\alpha$ are the identity matrix and Pauli-matrices acting on site $i$ respectively, and we will use this basis for our subsequent analysis. The proof of this statement is straightforward -- one notes that the reduced density matrix $\rho_A(t) = \Tr_B \ket{\psi(t)}\bra{\psi(t)}$ is in particular a Hermitian operator, and so must be a linear combination of $\mathcal{O}_A$ operators with real coefficients given by $ \frac{1}{2^{N_A}} \bra{\psi(t)} \mathcal{O}_A \ket{\psi(t)} = \frac{1}{2^{N_A}}  \bra{\psi} \mathcal{O}_A(t) \ket{\psi}$. Then, squaring $\rho_A(t)$ and using the orthonormality of $\mathcal{O}_A$ under the inner product, one obtains the claimed result. 

\section{Physical spreading of basis operators leads to entanglement growth}
We now use the above result to understand why entanglement typically grows in many-body systems. We consider the ensemble average of $S_2(t)$ over initial pure product states $\ket{\psi} = \ket{\vec{\sigma_1} \vec{\sigma}_2 \cdots \vec{\sigma}_N}$, where $\vec{\sigma}_i$ is the random direction that the $i$th spin is pointing to on its Bloch sphere.

It is convenient to switch from an average over an ensemble of initial states which are random product states to an average over an ensemble of locally-rotated Hamiltonians. A given initial product state can be written  as $\ket{\vec{\sigma_1} \vec{\sigma}_2 \cdots \vec{\sigma}_N} = \prod_i V_i \ket{\uparrow_1 \uparrow_2 \cdots \uparrow_N}$ where $V_i$ is the local unitary that rotates $\ket{\uparrow_i}$ into $\ket{\sigma_i}$. We can rewrite (\ref{eqn:measureOperators}) as
\begin{align}
\Tr_A(\rho_A^2(t)) &= \frac{1}{2^{N_A}} \sum_{\mathcal{O}_A} \bra{\uparrow \uparrow \uparrow} \tilde{U}_t^\dagger  \mathcal{V}^\dagger \mathcal{O}_A \mathcal{V} \tilde{U}_t \ket{\uparrow \uparrow \uparrow}^2 \nonumber \\
& = \frac{1}{2^{N_A}} \sum_{\mathcal{O}_A} \bra{\uparrow \uparrow \uparrow} \tilde{U}_t^\dagger   \mathcal{O}_A \tilde{U}_t \ket{\uparrow \uparrow \uparrow}^2,
\label{eqn:rotatedMeasurement}
\end{align}
where $\mathcal{V} \equiv \prod_i V_i $ and $\tilde{U}_t$ is the time evolution operator generated by the locally rotated Hamiltonian $\tilde{H} = \mathcal{V}^\dagger H \mathcal{V}$. This local rotation generates a new Hamiltonian $\tilde{H}$ that has the same locality properties as the original Hamiltonian $H$, i.e.~$\tilde{H} = \sum_X \tilde{H}_X$, which implies that both $H$ and $\tilde{H}$ have the same Lieb-Robinson (LR) velocities $v_{LR}$~\cite{LR1, LR2, LR3}, bounds which govern the speed of information and operator spreading in quantum many-body systems. The second equality arises from a straightforward statement of `basis invariance' of measurements~\cite{sup}. Thus, taking an ensemble average of $S_2(t)$ over initial product states in  (\ref{eqn:measureOperators}) is equivalent to taking an ensemble average over locally rotated Hamiltonians $\tilde{H}$ in (\ref{eqn:rotatedMeasurement}) with the measurement done in the particular state $\ket{\uparrow \uparrow \uparrow}$.

%Beginning from Eqn.~\ref{eqn:measureOperators},
%This last expression, Eqn.~\ref{eqn:S2time}, affords us a physical picture on how entanglement evolves as a function of time. If a basis operator $\mathcal{O}_A$ has support initially localized on a spatial region in $\Lambda$, then over time, it will spread in physical size to become a complicated larger operator, i.e. an operator with larger support than before. If the   density matrix $\rho(0)$ is `simple' enough (as in the case of an initial pure product state we will consider below), then we will gradually lose the ability to measure or resolve such operators. In that case, fewer and fewer terms contribute to the sum of Eqn.~\ref{eqn:S2time}, and so the entropy $S_2(t)$ increases, a fact we will make precise immediately below. 

We are now in a position to gain a physical understanding of how entanglement grows in a many-body system: under time evolution, a basis operator $\mathcal{O}_A$ with support in $A$ spreads on the lattice to become $\mathcal{O}_A(t) \equiv \tilde{U}_t^\dagger \mathcal{O}_A \tilde{U}_t$ (note the evolution under $\tilde{H}$), which has not only larger support but is also a complicated sum of other basis operators, though with conserved `weight', for example, \begin{align}
\sigma_i^x &\stackrel{t}{\to} \tilde{U}_t^\dagger \sigma_i^x \tilde{U}_t  =  c_i^x \sigma_i^x + c_{i,i+1}^{x,y} \sigma_i^x \sigma_{i+1}^y  + c_{i-1}^z \sigma_i^z \nonumber \\
&+ c_{i-1,i,i+1}^{z,x,z} \sigma_{i-1}^z \sigma_i^x \sigma_{i+1}^z +  \cdots \equiv \sum_{X,\mu} c_X^\mu \sigma_X^\mu,
\label{eqn:decomposition}
\end{align}
such that the total `weight' $\sum_{X,\mu} (c_X^\mu)^2 = 1$ for all times \cite{sup}. Then, the value of the measurement $\langle \uparrow \uparrow \uparrow | \mathcal{O}_A(t) | \uparrow \uparrow \uparrow \rangle$ typically decreases, as the complicated sum of operators will have many `off-diagonal' operators such as $\sigma_i^x \sigma_{i+1}^y$ etc.~that do not contribute, when only the `diagonal' operators which are products of $\sigma_i^z$s do, leading to an increase in entanglement entropy as calculated by (\ref{eqn:rotatedMeasurement}). Thus we see that entanglement growth in a many-body system is intimately related to the physical spreading of basis operators in real space: in some sense, EE increases because quantum information is `lost' in the inability of the state $\ket{\uparrow \uparrow \uparrow}$ to measure the increasingly complicated operator $\mathcal{O}_A(t)$. 

\section{A toy model for explaining universal linear growth in ergodic systems}
Equations (\ref{eqn:measureOperators}) and (\ref{eqn:decomposition}) are true regardless of the system in question, be it ergodic or otherwise. The differences in entanglement growth between different systems are completely captured in how an operator spreads and is decomposed in terms of other basis operators, c.f.~(\ref{eqn:decomposition}). Here, let us introduce a simple toy model where operators completely `scramble' in a linear lightcone, which gives a linear growth of entanglement. We believe that such a model captures the salient features of universal linear entanglement growth in generic \cite{Griffiths} ergodic many-body systems, as seen by Ref.~\cite{HuseFastEE}.

For local Hamiltonians with a bounded local Hilbert space, the velocity at which basis operators spreads can be at most linear, with an upper bound given by $v_{LR}$.  In other words, the operator can only spread within the LR light cone  $r \sim v_{LR}t$. The precise distribution of the coefficients $\{c_X^\mu\}$ in (\ref{eqn:decomposition}) depends on the Hamiltonian in question. In our toy model, we make a statistical statement about the distribution of coefficients. Let us assume that because of ergodicity, 1) a basis operator spreads linearly at some fixed velocity $v \leq v_{LR}$ (for all $\tilde{H}$) so that an initially local operator at $i$ spreads to become a sum of operators  contained within a ball of radius $(vt)^d$ centered at $i$, and 2) it is effectively `scrambled' in this ball. Precisely, we assume that an initially localized basis operator at $i$ has a decomposition as in (\ref{eqn:decomposition}) under time evolution, such that $\{c_X^\mu\}$  is a unit random vector of $4^{\gamma_d (vt)^d}$ coefficients, where $X$, the support of each basis operator in the decomposition is contained entirely within a ball of radius $(vt)^d$ centered at $i$. Such a picture is indeed supported by Ref.~\cite{LocalShock}, and we assume this behavior to be generically true for ergodic systems. We note that such a scrambling assumption, tied to ergodicity, will manifestly not hold for MBL systems: basis operators that have significant overlap with the local integrals of motions stay localized near site $i$, invalidating the assumption.

\begin{figure}[]
\center
\includegraphics[width=0.44\textwidth]{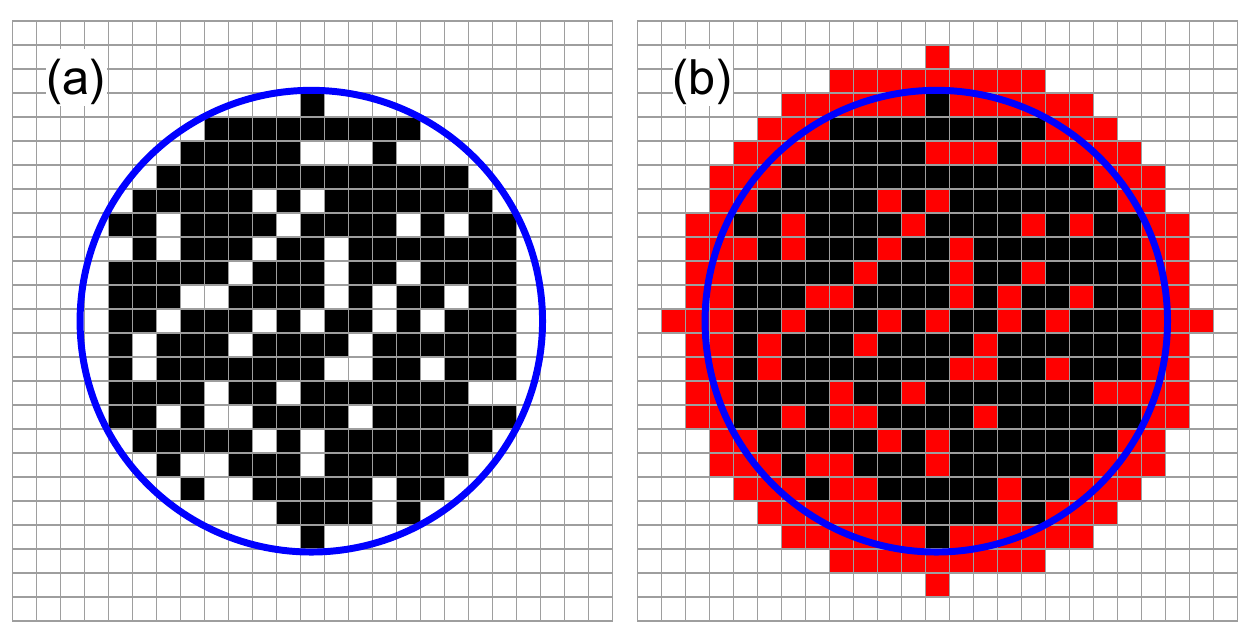}
\caption{(Color online) (a) The support of a typical basis operator $\mathcal{O}_A$. Here $\Lambda$ is a $d=2$ square lattice, and the subregion $A$ is a circle of radius $r_A$ demarcated by the blue circle. Black sites represent the presence of a non-trivial operator, $\sigma_i^x, \sigma_i^y$ or $\sigma_i^z$, while white sites represent the presence of an identity operator $\mathbb{I}_i$. One can see clusters of white sites within the circle. A site within a cluster is $\sim O(1)$ sites away from a black site. (b) Under time evolution, each black site spreads with velocity $v$, and so the clusters quickly get filled up (red sites) in time $vt \sim O(1)$, and simultaneously the operator also grows in physical size to become a ball of radius $r_A+vt$.}
\label{fig:typicalOp}
\end{figure}

Next, we estimate the measurement $\langle \uparrow \uparrow \uparrow | \mathcal{O}_{A}(t) | \uparrow \uparrow \uparrow \rangle$ in (\ref{eqn:rotatedMeasurement}). In principle, we  have to sum over all operators $\mathcal{O}_A$, but we can replace all $\mathcal{O}_A \to \mathcal{O}_{A,\text{typ.}}$, where $\mathcal{O}_{A,\text{typ.}}$ is a typical operator in subregion $A$ having support of size $\bar{L} = 3/4 N_A$. (Recall that the support of an operator are the sites where $\mathcal{O}_A$ acts non-trivially on, for example, $\mathcal{O}_A = \sigma_1^z \otimes \mathbb{I}_2 \otimes \sigma_3^x \otimes \mathbb{I}_4 \otimes \cdots \otimes \mathbb{I}_{N_A} $ has support on sites $1 \cup 3$ and size $2$.) The statement of the size being $\bar{L} = 3/4 N_A$ follows because in constructing some $\mathcal{O}_A$, there is a choice of three non-trivial operators $\{ \sigma_i^x, \sigma_i^y, \sigma_i^z \}$ compared with a choice of a single trivial operator $\mathbb{I}_i$ for every site, so there are ${ N_A \choose L} 3^L$ operators with support of size $L$, a skewed binomial distribution, which gives the average size $\bar{L} = 3/4 N_A$. Typicality follows because the standard deviation of the size of an operator $\sim \sqrt{N_A} \ll N_A$, coming from the same skewed binomial distribution. Thus, $\mathcal{O}_A(t)$ can be replaced by $\mathcal{O}_{A,\text{typ.}}(t)$ because the latter dominate the sum.

We now have to understand the time-evolution of $\mathcal{O}_{A,\text{typ.}}$.  Considered as an operator on the subregion $A$, $\mathcal{O}_{A,\text{typ.}}$ has $N_A - L$ sites on it acts trivially, i.e., with the identity operators.  We call groups of contiguous $\mathbb{I}_i$s `clusters'. A typical cluster has size $\sim O(1)$. Over time, each non-trivial single-site operator that makes up the typical operator  spreads, emitting a spherical `wavefront', and so the clusters quickly get filled up after $vt \sim O(1)$. After such time, the resulting operator will therefore have grown to become a complicated operator within a ball of radius $(r_A + vt)^d$, see Fig.~\ref{fig:typicalOp}. Following the previous discussion, we assume that the vector of coefficients $\{c_X^\mu\}$ denoting the decomposition of a typical time evolved operator into the basis operators of the ball is also a random vector but now with dimension $4^{\gamma_d (r_A+vt)^d}$.

Next, we estimate the measurement of a typical operator. This is given by
\begin{align}
\langle \uparrow \uparrow \uparrow | \mathcal{O}_{A,\text{typ.}}(t) | \uparrow \uparrow \uparrow \rangle^2 = \frac{2^{\gamma_d(r_A+vt)^d} }{4^{\gamma_d(r_A+vt)^d}},
\label{eqn:typ}
\end{align}
which can be understood as follows. In the decomposition $\mathcal{O}_{A,\text{typ.}}(t) = \tilde{U}_t^\dagger \mathcal{O}_{A,\text{typ.}} \tilde{U}_t = \sum_\mathcal{O} c_\mathcal{O}(t) \mathcal{O}$, the only operators $\mathcal{O}$ which give a non-zero overlap with $ |\uparrow \uparrow \uparrow \rangle$ are the `diagonal' operators $\mathbb{I}, \sigma_i^z, \sigma_i^z \sigma_j^z, \cdots$. There are $2^{\gamma_d (r_A+vt)^d}$ such operators contained in a ball of radius $(r_A+vt)^d$. The coefficient $c_\mathcal{O}$ that accompanies such an operator is a random variable (which follows from our scrambling assumption) and has typical magnitude $1/\sqrt{\text{dim(Ball)}(t)} = 1/\sqrt{4^{\gamma_d (r_A+vt)^d}}$. Furthermore, since the coefficients $c_\mathcal{O}(t)$ are independent random variables for different operators $\mathcal{O}$, the cross-terms in the LHS of (\ref{eqn:typ}), $\Big(\sum_{\mathcal{O} } c_\mathcal{O}(t) \Big)^2 $, vanish, reducing the sum to $\sum_{\mathcal{O} } c_\mathcal{O}(t)^2 = \frac{2^{\gamma_d(r_A+vt)^d} }{4^{\gamma_d(r_A+vt)^d}}$, the RHS. 

Finally, we arrive at the expression for $S_2(t)$ and obtain the universal linear growth. Plugging the above result into (\ref{eqn:rotatedMeasurement}) and simplifying, we get
\begin{align}
\text{Tr}_A(\rho_A^2(t)) = 2^{-\gamma_d((r_A+vt)^d - r_A^d)}.
\end{align}
This calculation is valid until $vt \lesssim (2^{1/d}-1) r_A$, at which point $\Tr_{A}(\rho_A^2(t)) \sim 1/2^{N_A}$. Then, the error from neglecting the contribution of the global identity term $\mathbb{I}_{\mathcal{H}_A}$ becomes non-negligible, since the identity operator never spreads under time evolution and will always contribute a factor of $1/2^{N_A}$. Thus, the EE saturates to the maximum allowed for region $A$. The second R\'enyi EE is thus
\begin{align}
S_2(t) = \begin{cases}
 \gamma_d ((r_A + vt)^d - r_A^d) \log 2 & \text{ for } \frac{\mathcal{O}(1)}{v} < t \lesssim \frac{f_d r_A}{v}, \\
 \gamma_d r_A^d \log 2 = N_A \log 2 & \text{ for } t \gtrsim \frac{f_d r_A}{v},
\end{cases}
\end{align}
where $f_d  = (2^{1/d}-1)$. For times $t < 2r_A/v(d-1)$, $S_2(t)$ can be approximated by a linear function
\begin{align}
S_2(t) \sim  \left[ (\gamma_d  d  r_A^{d-1})  v  t \right] \log 2.
\label{eqn:EEgrowth}
\end{align}
In fact, this linear approximation is always valid as long as the $S_2(t)$ calculation is valid, since $2r_A/v(d-1) > f_d r_A/v$ for all $d$. We see that the expression for the dynamics of entanglement, (\ref{eqn:EEgrowth}), has three terms: 1) a geometric factor $\gamma_d d r_A^{d-1}$ which gives the area of an entanglement `wavefront', 2) the speed of the wavefront $v$, and 3) a universal linear time dependence $t$. Thus this has the interpretation of entanglement spreading in a `tsunami', similar to a view put forth in Ref.~\cite{HolographicGrowth1}.

%To sum up, assuming Eqn.~\ref{eqn:measureOperators}, we showed that the linear in time growth of entanglement in ergodic many-body systems is universal, with a prefactor that is non-universal which depends on the geometry of $A$ and also the rate of spreading of local operators $v$, set by the parameters of the Hamiltonian such as the interaction strength.

\begin{comment}
 we see that $\Tr(\rho_A^2(t))$ can also be evaluated by measuring all time evolved-operators $T_{i_1, \cdots i_L}^{a_{i_1}, \cdots, a_{i_L}}(t) = U_t^\dagger \left( T_{i_1, \cdots i_L}^{a_{i_1}, \cdots, a_{i_L}} \right) U_t$.
\end{comment}

\section{Proposal to experimentally measure $S_n(t)$ and relation to Loschmidt echo} 
Here we propose a way to experimentally measure $S_n(t)$ via a local measurement. We consider an replicated system consisting of $n$ disjoint copies of the original system, each with a similar bipartition $A_i$ and $B_i$. We arrange them in a way similar to Ref.~\cite{AbaninEE}: a star geometry such that the boundaries between subsystems $A_i$ and $B_i$ are placed near each other, see Fig.~\ref{fig:QS}.  We then introduce a quantum switch (a 2-level system), which if in the $\ket{\!\uparrow }$ state, allows tunneling only between subregions $A_i$ and $B_i$ such that the full Hamiltonian on the composite system is $H$, while if in the $|\!\downarrow \rangle$ state, allows tunneling only between subregions $A_i$ and $B_{i+1}$, such that the full Hamiltonian is $H+V$. Here $H = \sum_{i=1}^n H_i = \sum_{i=1}^n H_{A_i}+ H_{B_i} + H_{A_i B_i}$, where each $H_i$ is a copy of the Hamiltonian, (\ref{eqn:H}), acting on the $i$th-copy of the Hilbert space $\mathcal{H}_i = \mathcal{H}_{A_i} \otimes \mathcal{H}_{B_i}$, and $V$ the local `reconnection' operator, $V = \sum_{i=1}^n H_{A_i B_{i+1}} - H_{A_i B_i}$, with the cyclic condition $n+1 = 1$, so that $H+V = \sum_{i=1}^n H_{A_i}+ H_{B_i} + H_{A_i B_{i+1}}$. If we prepare the state of the replicated system as $\ket{\psi^n } \equiv \otimes_{i=1}^n \ket{\psi}_i $ where $\ket{\psi}$ is the initial product state of the original system and also the quantum switch as a maximally entangled Bell state, so that the full state is $\ket{\Psi} = \ket{\psi^n} \otimes  \frac{1}{\sqrt{2}} ( \ket{\!\!\uparrow} + \ket{\!\!\downarrow} )$, then under unitary time evolution, 
\begin{align}
\ket{\Psi(t)} =  \frac{1}{\sqrt{2}} \left(e^{-iHt}  \ket{\psi^n \!   \uparrow} + e^{-i(H+V)t}  \ket{\psi^n \!\downarrow} \right).
\end{align}
Measuring $\sigma^x(t)$ of the quantum switch, a local measurement, gives
\begin{align}
\text{Tr}_A(\rho_A^n(t)) = \langle \psi^n | e^{i(H+V)t} e^{-iHt} | \psi^n \rangle \equiv \mathcal{F}(t),
\label{eqn:F}
\end{align}
from which the $n$th R\'enyi entropy can be obtained: $S_n(t) = \frac{1}{1-n} \log\mathcal{F}(t)$. Thus we see that the dynamics of entanglement is indeed an observable local quantity if we work in an extended system. See also Refs.~\cite{EEMeasurement1, EEMeasurement2} on other theoretical proposals to measure EE and its growth. Additionally, Ref.~\cite{Islam2015} has experimentally measured EE in a Bose-Hubbard system also in a replicated set-up. However, the difference of our proposal with their experimental technique is that we only require a single {\it local} measurement wherelse \cite{Islam2015} requires an extensive number of measurements (namely, measuring the parity on all sites).

The above claim is based on an alternative reformulation of $S_n(t)$, relating it to a Loschmidt echo $\mathcal{F}(t)$~\cite{LE4, LE2, LE3, LE1, LE5, LE6} on the replicated system.  It is a well-known trick \cite{sup, EEMeasurement1, EETrick}  that the trace of the $n$th power of $\rho_A$ can be calculated by replicating $n$ copies of the original Hilbert space $\mathcal{H} = \mathcal{H}_A \otimes \mathcal{H}_B$, and calculating the swap operator $S$ that cyclically swaps subregions $A_i \to A_{i \text{mod} n + 1}$: if $\Omega = \rho^{\otimes n}$ is the $n$ replicated density matrix, then
$
\Tr_{A}(\rho_A^n) = \Tr_{\mathcal{H}^{\otimes n}}( \Omega S).
$

Now let us specialize to the case considered before: when the initial state on one system is a random product state $\ket{\psi} = \ket{\vec{\sigma}_1 \vec{\sigma}_2\cdots \vec{\sigma}_N}$ and  $\rho = \ket{\psi}\bra{\psi}$. If each $\rho$ evolves independently under the Hamiltonian given by (\ref{eqn:H}), then $\Omega$ evolves under the composite Hamiltonian $H = \sum_{i=1}^n H_i$. Then, 
\begin{align}
\Tr_{A}(\rho_A^n(t)) =  \bra{ \psi^n} e^{i H t} S e^{-i Ht} \ket{\psi^n}.
\end{align}
However, because the initial state is a product state, $\bra{\psi^n}S^\dagger = \bra{\psi^n}$, and since $S^\dagger e^{i H t} S  = e^{i (H+V) t}$, we end up with the claimed result, (\ref{eqn:F}), which is a Loschmidt echo $\mathcal{F}(t)$ with `perturbation' $V$ to $H$. Note that an analogous statement also holds for time-dependent Hamiltonians even though the unitary time evolution operator will not have the form $e^{-iHt}$.

\section{Conclusion}
 We have provided a theoretical description and a physical picture of the entanglement dynamics in a many-body system by relating entanglement growth to the physical spreading of basis operators on a lattice. We  also introduced a simple toy model where an initial basis operator effectively scrambles in its light cone, thus producing a linear entanglement growth, which we believe captures the salient features of the universal linear in time growth of entanglement seen in ergodic many-body systems. Furthermore, the entanglement growth in this model has an interpretation of an `entanglement tsunami': there is a prefactor which depends on the geometry of the subregion $A$, which propagates outwards in a `wavefront' at  speed $v$. We remark that while entanglement growth in MBL systems has been successfully explained through a local integrals of motion picture, it is an interesting question to explain this dynamics in terms of our operator spreading language.

We have also provided an alternative interpretation of the growth of the $n$-th R\'enyi entropies as a Loschmidt echo in a composite system, subject to a perturbation that reconnects different subregions. By using this reformulation, we have proposed an experimental way of measuring the growth of the R\'enyi entropies: one can effect the reconnection by a quantum switch and measure the state of the quantum switch to extract the Loschmidt echo and hence the EE. This proposal can be used, for example, to directly detect MBL phases by measuring the logarithmic growth of entanglement. 

\textit{Note added.} -- Recent works \cite{2016arXiv160805101M, 2016arXiv160806950N} also studied entanglement dynamics in chaotic many-body systems, and argued that entanglement growth speed is bounded from above by the operator spreading speed.  In our operator counting language picture, this can be accounted for in our toy model by refining the `scrambling' assumption which gives rise to the measurement value (\ref{eqn:typ}): there will be a distribution of coefficients that presumably depends upon the size of the operators in the decomposition (\ref{eqn:decomposition}), which we leave for future work to explore.

%%%

%\textit{Acknowledgments.} -- This work was partially supported by Sloan Foundation, Ontario Early Researcher Award and NSERC Discovery Grant.  

%\bibliography{refs}
\bibliography{EntanglementDynamicsv8PRB}

\begin{thebibliography}{35}
\bibitem [{Gri()}]{Griffiths}%
  \BibitemOpen
  \href@noop {} {}\bibinfo {note} {By generic, we exclude delocalized systems
  with rare Griffiths regions. The presence of these regions causes the
  entanglement growth to be sublinear \cite{PhysRevB.93.060201}, for which our
  toy model would be inapplicable. This is perfectly consistent with our
  `operator spreading picture': in such Griffiths regions, transport and hence
  operator spreading is suppressed and so the complete scrambling within a
  linear lightcone assumption would not hold; a modified model can be
  formulated to account for these effects giving rise to a different form for
  entanglement growth.}\BibitemShut {Stop}%
\end{thebibliography}

\appendix

\section{Appendix A: Analysis for systems with bounded Hilbert spaces of other dimensions}
Here we consider the extension of our analysis and results for systems with bounded Hilbert spaces of other dimensions, $\dim(\mathcal{H}_i) = k < \infty$, so that the full Hilbert space is $\mathcal{H} = \otimes_{i=1}^N \mathcal{H}_i$.

\subsection{I. Results}
We simply state the corresponding results here. We always work with a random pure product state $\ket{\psi}$ as the initial state, and $\rho = \ket{\psi}\bra{\psi}$. The second R\'enyi entropy $S_2(t) = -\log \text{Tr}(\rho^2_A(t))$, like in the discussion of the main text for spin-$1/2$ systems, is  similarly related to the measurement of all independent operators with support in $A$:
\begin{align}
\text{Tr}_A(\rho_A^2(t)) = \frac{1}{k^{N_A}} \sum_{ \mathcal{O}_A} \Big( \Tr( \rho \mathcal{O}_A(t)) \Big)^2,
\label{eqn:measureOperators2}
\end{align}
where  $N_A$ is the total number of sites in $A$. The operators $\mathcal{O}_A$ are assumed to be Hermitian and independent, as defined by the Hilbert-Schmidt inner product on operators: $\frac{1}{k^{N_A}} \text{Tr}_A(\mathcal{O}_A^\dagger \mathcal{O}'_A)= \delta_{\mathcal{O}_A, \mathcal{O}'_A}$, and they form a basis for all Hermitian operators on $A$ (see Appendix A.II).

Following the same logic as in the main text, assuming that each basis operator $\mathcal{O}_A$ effectively scrambles in its lightcone $vt$, we end up with:
\begin{align}
\text{Tr}_A(\rho_A^2(t)) = k^{-\gamma_d((r_A+vt)^d - r_A^d)},
\label{eqn:Growth}
\end{align}
valid until $vt \lesssim (2^{1/d}-1) r_A$, at which point $\Tr_{A}(\rho_A^2) \sim 1/2^{k_A}$. Then
\begin{align}
S_2(t) = \begin{cases}
 \gamma_d ((r_A + vt)^d - r_A^d) \log k & \text{ for } \frac{\mathcal{O}(1)}{v} < t \lesssim \frac{f_d r_A}{v}, \\
 \gamma_d r_A^d \log k = N_A \log k & \text{ for } t \gtrsim \frac{f_d r_A}{v},
\end{cases}
\end{align}
where $f_d  = (2^{1/d}-1)$. For times $t < 2r_A/v(d-1)$, $S_2(t)$ is approximated by a linear function
\begin{align}
S_2(t) \sim  \left( (\gamma_d  d  r_A^{d-1}) \times v \times  t \right) \log k
\end{align}
which is always valid as long as the $S_2(t)$ calculation is valid, since $2r_A/v(d-1) > f_d r_A/v$ for all $d$.  We see that the universal linear growth of EE is still present for systems with different $\dim(\mathcal{H}_i) = k < \infty$.

\subsection{II. Derivation of results}
Here we perform two things. First, we prove (\ref{eqn:measureOperators2}), that is,
\begin{align}
\text{Tr}_A(\rho_A^2(t)) = \frac{1}{k^{N_A}} \sum_{ \mathcal{O}_A} \Big( \Tr( \rho \mathcal{O}_A(t)) \Big)^2,
\end{align}
and second, we prove (\ref{eqn:Growth}), 
\begin{align}
\text{Tr}_A(\rho_A^2(t)) = k^{-\gamma_d((r_A+vt)^d - r_A^d)},
\end{align}
repeating the arguments presented in the main text for a system with $\dim(\mathcal{H}_i) = k < \infty$.

\subsubsection{Proving (\ref{eqn:measureOperators2})}

Let $\dim(\mathcal{H}_i) = k < \infty$. Then, a basis for all Hermitian operators acting on a given site $i$ is given by the traceless, Hermitian generators $T_i^a$, ($a = 1,\cdots, k^2-1$), of the fundamental representation of the Lie algebra $su(k)$, together with the identity operator $\mathbb{I}_{i,k} \equiv T_i^0$. Note that the space of all Hermitian operators on site $i$ is itself a Hilbert space $\mathcal{H}^O_i$ of dimensionality $k^2$, which should not be confused with the Hilbert space of quantum states $\mathcal{H}_i$. The generators $T_i^a$ ($a \neq 0$) can be chosen to obey the relations
\begin{align}
T_i^a T_i^b = \delta_{ab} \mathbb{I}_{i,k} + \sqrt{\frac{k}{2}} \sum_{c = 1}^{k^2-1} (i f_{abc} + d_{abc}) T_i^c,
\label{eqn:Tproduct}
\end{align}
where $f$ are the antisymmetric structure constants and $d$ the symmetric ones. Under the Hilbert-Schmidt inner product, the above operators are thus orthogonal:
\begin{align}
\frac{1}{k}\Tr_{\mathcal{H}_i}(T_i^{a \dagger} T_i^b) = \frac{1}{k} \delta_{ab} \Tr_{\mathcal{H}_i}(\mathbb{I}_{i,k}) = \delta_{ab}.
\label{eqn:HS}
\end{align}
For example, for the case of $k = 2$, the standard basis operators on site $i$ can be taken to be the familiar Pauli matrices: $\{\mathbb{I}_{i}, \sigma_i^x, \sigma_i^y, \sigma_i^z \}$.

To define a basis on the space of Hermitian operators on $N$ sites $\mathcal{H}^O = \otimes_i \mathcal{H}_i^O$, we form tensor products of $T_i^a$s, so that a basis operator 
\begin{align}
\mathcal{O} \in \otimes_{i=1}^N \{\mathbb{I}_i, T_i^1, \cdots, T_i^{k^2-1} \}.
\end{align}
There are $(k^2)^{N}$ such basis operators in total, and they are orthonormal as given by the generalized Hilbert-Schmidt inner product:
\begin{align}
\frac{1}{k^N} \text{Tr}(\mathcal{O}^\dagger  \mathcal{O}' ) = \frac{1}{k^N} \text{Tr}(\mathcal{O} \mathcal{O}' )= \delta_{ \mathcal{O},  \mathcal{O}'}.
\label{eqn:HS}
\end{align}
An example of a basis operator is $\mathcal{O} \equiv  T_{i_1, \cdots i_L}^{a_{i_1}, \cdots, a_{i_L}} = T_{i_1}^{a_{i_1}} \otimes T_{i_2}^{a_{i_2}} \otimes \cdots \otimes T_{i_L}^{a_{i_L}} \otimes \mathbb{I}_{\Lambda - i_1 \cup \cdots \cup i_L}$ which is a product of $T_{i}^a$ operators on $L$ distinct sites and a global identity acting on the subregion $\Lambda - i_1 \cup \cdots \cup i_L$ ($\Lambda$ is the full system). We say that such an operator has (physical) size $L$ and has support on subregion $i_1 \cup \cdots \cup i_L$.

Since the reduced density matrix is also a Hermitian operator, it can be written as
%\begin{align}
%\rho = \frac{1}{k^N} \mathbb{I}_{\mathcal{H}} + \sum_{\stackrel{L=1}{\{i_1, \cdots i_L\}}}^N c_{i_1, \cdots i_L}^{a_{i_1}, \cdots, a_{i_L}} T_{i_1, \cdots i_L}^{a_{i_1}, \cdots, a_{i_L}},
%\end{align}
\begin{align}
\rho_A(t) = \sum_{\mathcal{O}_A} c_{\mathcal{O}_A}(t) \mathcal{O}_A,
\end{align}
where the  coefficient $c_\mathcal{O}$ is a real number given by
%\begin{align}
%c_{i_1, \cdots i_L}^{a_{i_1}, \cdots, a_{i_L}}  = \frac{1}{k^N} \Tr_{\mathcal{H}}\left[ \rho T_{i_1, \cdots i_L}^{a_{i_1}, \cdots, a_{i_L}}\right].
%\label{eqn:coeff}
%\end{align}
\begin{align}
c_{\mathcal{O}_A}(t) = \frac{1}{k^{N_A}} \text{Tr}(\rho {\mathcal{O}_A(t)}).
\end{align}
Squaring it gives
\begin{align}
\rho_A^2(t) = \sum_{\mathcal{O}_A,  \mathcal{O}'_A} c_{\mathcal{O}_A}(t) c_{\mathcal{O}'_A}(t) \mathcal{O}_A  \mathcal{O}'_A,
\end{align}
and making use of the Hilbert-Schmidt inner product, (\ref{eqn:HS}), we get
\begin{align}
\text{Tr}_A(\rho_A^2(t)) = k^{N_A} \sum_{\mathcal{O}_A} c_{\mathcal{O}_A}(t)^2 = \frac{1}{k^{N_A}} \sum_{\mathcal{O}_A}\Big( \text{Tr}(\rho \mathcal{O}_A(t)) \Big)^2,
\end{align}
our desired result. Thus, we see that second R\'enyi EE $S_2(t) = -\log \text{Tr}_A(\rho_A^2(t))  $ can be obtained by measuring, witin the fixed initial state $\rho$, all time evolving basis operators $\mathcal{O}_A(t)$ that have support initially in $A$.

\subsubsection{Proving (\ref{eqn:Growth})}
We want to prove
\begin{align}
\text{Tr}_A(\rho_A^2(t)) = k^{-\gamma_d((r_A+vt)^d - r_A^d)},
\end{align}
once again for a subregion $A$ which is a $d$-ball of volume $\gamma_d r_A^d = N_A$.

Given a random initial product state, and following the main text, we can write it as 
\begin{align}
\ket{\phi_1 \phi_2 \cdots \phi_N} = \prod_i V_i  \ket{1_1 1_2 \cdots i_N}
\end{align}
%\begin{align}
%\rho(0) &= \bigotimes_{i=1}^N \left( V_i \ket{1_i}\bra{1_i} V_i^\dagger\right)  = \prod_i V_i \ket{1 }\bra{1} \prod_i V_i^\dagger,
%\label{eqn:rhoInitial}
%\end{align}
where $\ket{1_i}$ is the first basis vector in $\mathcal{H}_i$ for some enumeration of the $k$ basis vectors (similar to $\ket{\!\uparrow_i}$ in the case of $SU(2)$) and $V_i$ the $SU(k)$ operator (a local unitary operator acting only on site $i$) that rotates $\ket{1_i}$ into $\ket{\phi_i}$. Let us call $\ket{1} \equiv \ket{1_1 1_2 \cdots 1_N}$. Then (\ref{eqn:measureOperators2}) becomes
\begin{align}
\Tr_{A}(\rho_A^2(t)) = \frac{1}{k^{N_A}} \sum_{\mathcal{O}_A} \bra{1} \tilde{U}_t^\dagger \mathcal{O}_A \tilde{U}_t \ket{1}^2,
\label{eqn:generalizedS2}
\end{align}
where the unitary time evolution operator is now generated by the locally rotated Hamiltonian $ \prod_i V_i^\dagger  (H) \prod_i V_i $, which is still a local Hamiltonian having the same lightcone velocity as $H$.

Now, we want to replace all terms $\bra{1} \tilde{U}_t^\dagger \mathcal{O}_{A} \tilde{U}_t \ket{1}^2$ in the sum of (\ref{eqn:generalizedS2}) by the contribution due to a typical operator $\bra{1} \tilde{U}_t^\dagger \mathcal{O}_{A,\text{typ.}} \tilde{U}_t \ket{1}^2$. Following the main text, the average size/support of an operator in the set of basis operators of $A$ is $\bar{L} = (k^2-1)/k^2 \times N_A$, which follows from a skewed binomial distribution as there are $k^2-1$ choices of non-trivial operators $T_i^a$ acting on a site, compared to a choice of a single local identity $\mathbb{I}_i$. In fact, an average operator is also typical, because the standard deviation scales as $\sqrt{N_A}$, which is smaller than $N_A$. Thus the error induced by replacing all contributions in the sum with that of a typical operator is small because the sum is dominated by the contributions from typical operators anyway.

Next, we invoke the same `scrambling' assumption for ergodic systems. Assume that a basis operator localized at site $i$ scrambles in a ball of radius $(vt)^d$; then, a typical operator of mean size $\bar{L} = (k^2-1)/k^2 \times N_A$ becomes a complicated operator in a ball of radius $(r_A + vt)^d$. Precisely, the vector of $(k^2)^{\gamma_d(r_A+vt)^d}$ coefficients  $c_\mathcal{O}(t)$ that denotes the decomposition of a time-evolved operator $\mathcal{O}_A(t) = \sum_{\mathcal{O}} c_\mathcal{O}(t) \mathcal{O}$ in terms of basis operator of the ball is a random vector, after time $vt \sim \mathcal{O}(1)$. 

To estimate $\bra{1} \tilde{U}_t^\dagger \mathcal{O}_{A,\text{typ.}} \tilde{U}_t \ket{1}^2$, we have to count how many `diagonal' operators $\mathcal{O}_{A,\text{diag.}}$ there are in the set of operators that span the ball of radius $(r_A+vt)^d$, such that $\bra{1} \mathcal{O}_{A,\text{diag.}} \ket{1} = 1$ and $\bra{1} \mathcal{O}_{A,\text{not diag.}} \ket{1} = 0$. These diagonal basis operators are in fact, the given by products of $T_i^a$ which live in the Cartan sub-algebra of $su(k)$ (i.e. the diagonal generators), together with the identity operator on a site $\mathbb{I}_i$. For a given site, there are $k$ possible choices only. Thus the total number of diagonal basis operators in the ball of radius $(r_A+vt)^d$ is $k^{\gamma_d (r_A+vt)^d}$. For the familiar case of $su(2)$, i.e.~a spin-$1/2$ chain, there is only $1$ generator in the Cartan sub-algebra, $\sigma^z_i$, so together with the identity operator, this implies that the number of diagonal operators in the ball are $2^{\gamma_d (r_A+vt)^d}$, as quoted in the main text.

Then, following the same logic as the main text, because the coefficients $c_\mathcal{O}$ in the decomposition of a time-evolved basis operator $\mathcal{O}_A(t)$ are independent random variables, by assumption, the cross-terms $c_\mathcal{O} c_{\mathcal{O}'}$ that arise in $\bra{1} \tilde{U}_t^\dagger \mathcal{O}_{A,\text{typ.}} \tilde{U}_t \ket{1}^2$ do not contribute, and so 
\begin{align}
\bra{1} \tilde{U}_t^\dagger \mathcal{O}_{A,\text{typ.}} \tilde{U}_t \ket{1}^2 = \sum_{\mathcal{O}_{\text{diag.}}} c_{\mathcal{O}_{\text{diag.}}}^2(t) = \frac{k^{\gamma_d (r_A+vt)^d}}{(k^2)^{\gamma_d (r_A+vt)^d}}.
\end{align}
Thus, we end up with
\begin{align}
\text{Tr}_A(\rho_A^2(t)) &\approxeq \frac{1}{k^{N_A}} \sum_{\mathcal{O}_A} \bra{1} \tilde{U}_t^\dagger \mathcal{O}_{A,\text{typ.}} \tilde{U}_t \ket{1}^2 \nonumber \\
& = \frac{1}{k^{\gamma_d r_A^d}} (k^2)^{\gamma_d r_A^d} \frac{k^{\gamma_d (r_A+vt)^d}}{(k^2)^{\gamma_d (r_A+vt)^d}} \nonumber \\
& = k^{-\gamma_d ( (r_A+vt)^d - r_A^d )},
\end{align}
the desired result.

\section{Appendix B: Basis invariance for measurement}
Let $\dim(\mathcal{H}_i) = k < \infty$. Given any Hermitian operator $H$, (which includes $\rho$, the density matrix), and a bipartition of the system into $A$ and $B$, we prove that the 	`weight' of the operator in $A$, given by
\begin{align}
\text{Weight} = \sum_{\mathcal{O}_A} \Big( \text{Tr}(H \mathcal{O}_A) \Big)^2,
\end{align}
is independent of the orthonormal basis set, $\{ \mathcal{O}_A \}$, that spans all Hermitian operators in $A$. The inner product on this set is given by the Hilbert-Schmidt norm: $\frac{1}{k^{N_A}} \text{Tr}_A(\mathcal{O}_A^\dagger \mathcal{O}'_A) = \delta_{\mathcal{O}_A \mathcal{O}'_A}$.

\subsection{I: Proof}

The above expression can be written as $\sum_{\mathcal{O}_A} \text{Tr}_A( H_A \mathcal{O}_A)^2$, where  $H_A = \text{Tr}_B H$.

Now, $H_A$ can be decomposed as
\begin{align}
H_A = \sum_{\mathcal{O}_A} h_{\mathcal{O}_A} \mathcal{O}_A.
\end{align}
Then,
\begin{align}
\text{Tr}_A(H_A^2) & = \text{Tr}_A \left( \sum_{\mathcal{O}_A \mathcal{O}'_A } h_{\mathcal{O}_A}h_{\mathcal{O}'_A} \mathcal{O}_A \mathcal{O}'_A \right) \nonumber \\
& = \sum_{\mathcal{O}_A \mathcal{O}'_A } h_{\mathcal{O}_A}h_{\mathcal{O}'_A} \text{Tr}_A( \mathcal{O}_A \mathcal{O}'_A) \nonumber \\
& = \sum_{\mathcal{O}_A \mathcal{O}'_A } h_{\mathcal{O}_A}h_{\mathcal{O}'_A} k^{N_A} \delta_{\mathcal{O}_A \mathcal{O}'_A} \nonumber \\
& = k^{N_A} \sum_{\mathcal{O}_A} h_{\mathcal{O}_A}^2 \nonumber \\
& = k^{N_A} \sum_{\mathcal{O}_A} \left( \frac{\text{Tr}_A( H \mathcal{O}_A)}{k^{N_A}} \right)^2 \nonumber \\
& = \frac{1}{k^{N_A}} \sum_{\mathcal{O}_A} \Big( \text{Tr}_A  (H \mathcal{O}_A) \Big)^2.
\end{align}
We have used $\mathcal{O}_A = \mathcal{O}_A^\dagger$, and also the orthonormality of the basis set under the Hilbert-Schmidt norm in the above.

Since the LHS is independent of the basis set used, the RHS must be as well, and the weight is given by $k^{N_A} \text{Tr}_A (H_A^2)$. 

\subsection{II: Proof of conversation of weight of an operator under time evolution}
The conservation of weight of any Hermitian operator $H$ under time evolution $H(t) = U_t H U_t^\dagger$ is a direct application of the above statement, with the subregion $A$ being the full space.  

The weight of a time-evolving operator $H(t) = \sum h_{\mathcal{O}}(t) \mathcal{O}$,
given by the sum 
\begin{align}
\sum_{\mathcal{O}} h_\mathcal{O}(t)^2,
\end{align}
is conserved, since $h_\mathcal{O}(t) = \frac{1}{k^{N_A}} \text{Tr}(U_t H U_t^\dagger \mathcal{O} ) = \frac{1}{k^{N_A}} \text{Tr}( H U_t^\dagger \mathcal{O} U_t) $, and the set $\{ U_t^\dagger \mathcal{O} U_t \}$ is an orthonormal basis, if $\{ \mathcal{O} \}$ is, at any time $t$. Thus, the weight is constant in time and proportional to $\text{Tr}(H^2)$.

\section{Appendix C: Proof of replica trick}
We prove the statement
\begin{align}
\text{Tr}_A(\rho_A^n) = \text{Tr}_{\mathcal{H}^{\otimes n} } ( \Omega S),
\end{align}
where $\mathcal{H}^{\otimes n} = \bigotimes_{i=1}^n \mathcal{H}_{A_i} \otimes \mathcal{H}_{B_i}$ is an extended, composite Hilbert space comprised of $n$ copies of the original Hilbert space $\mathcal{H}$ with a bipartition into $A$ and $B$. Here $\Omega = \rho^{\otimes n}$, the $n$-replicated density matrix, where $\rho$ is the density matrix on a single copy of the Hilbert space. $S$ is the unitary swap operator that cyclically permutes only states of $A_i$:
\begin{align}
& S \ket{\psi_1}_{A_1} \ket{\psi_2}_{A_2}\cdots \ket{\psi_n}_{A_n} \otimes \ket{\phi}_{B_1,\cdots,B_n} \nonumber \\
&= 
 \ket{\psi_n}_{A_1} \ket{\psi_1}_{A_2} \cdots \ket{\psi_{n-1}}_{A_n} \otimes \ket{\phi}_{B_1,\cdots, B_n}.
\end{align}
We start from the RHS, $\text{Tr}_{\mathcal{H}^{\otimes n} } ( \Omega S)$. Let $\ket{\psi_{a_1} \cdots \psi_{a_n} \phi_{b_1} \cdots \phi_{b_n}} = \ket{\psi_{a_1}}_{A_1} \otimes \cdots \otimes \ket{\psi_{a_n}}_{A_n} \otimes \ket{\phi_{b_1}}_{B_1} \otimes \cdots \otimes \ket{\phi_{b_n}}_{B_n}$, where $\ket{\psi_{a_i}}_{A_i}$ is a set of states labeled by $a_i$ which spans $\mathcal{H}_{A_i}$, and similarly $\ket{\phi_{b_i}}$ is a set of states labeled by $b_i$ which spans $\mathcal{H}_{B_i}$. Then
\begin{align}
&\text{Tr}_{\mathcal{H}^{\otimes n} } ( \Omega S) \nonumber  \\
&= \sum_{\stackrel{a_1, \cdots, a_n}{ b_1, \cdots, b_n}} \bra{\psi_{a_1} \cdots \psi_{a_n} \phi_{b_1} \cdots \phi_{b_n}} \Omega S  \ket{\psi_{a_1} \cdots \psi_{a_n} \phi_{b_1} \cdots \phi_{b_n}} \nonumber \\
& = \sum_{\stackrel{a_1, \cdots, a_n}{ b_1, \cdots, b_n}} \bra{\psi_{a_1} \cdots \psi_{a_n} \phi_{b_1} \cdots \phi_{b_n}} \Omega \ket{\psi_{a_n} \cdots \psi_{a_{n-1}} \phi_{b_1} \cdots \phi_{b_n}}  \nonumber \\
& \equiv \sum_{ a_1, \cdots, a_n} \bra{\psi_{a_1} \cdots \psi_{a_n}} \left( \text{Tr}_B \rho  \right)^{\otimes n}  \ket{\psi_{a_n} \cdots \psi_{a_{n-1}}}  \nonumber \\
& = \sum_{ a_1, \cdots, a_n} \bra{\psi_{a_1}} \rho_A \ket{\psi_{a_n}} \bra{\psi_{a_2}} \rho_A \ket{\psi_{a_1}} \cdots \bra{\psi_{a_n}} \rho_A \ket{\psi_{a_{n-1}}} \nonumber \\
& =  \sum_{ a_1, \cdots, a_n} \bra{\psi_{a_1}} \rho_A \ket{\psi_{a_n}} \bra{\psi_{a_n}} \rho_A \ket{\psi_{a_{n-1}}} \cdots \bra{\psi_{a_2}} \rho_A \ket{\psi_{a_{1}}} \nonumber \\
& = \sum_{a_1} \bra{\psi_{a_1}} \rho_A^n \ket{\psi_{a_1}} \nonumber \\
& \equiv \text{Tr}_A(\rho_A^n),
\end{align}
which is the desired result. In the above, we have used the resolution of the identity $\mathbb{I}_{\mathcal{H}_{A}} = \sum_{a} \ket{\psi_{a}} \bra{\psi_{a}}$.

\end{document}